\documentclass[twocolumn,english]{article}
\usepackage{spconf,graphicx,amsmath,subeqnar,overpic}
\usepackage{amsmath,amsfonts,amscd,amssymb}
\usepackage[justification=justified]{caption}
\usepackage{color}


\title{The control structure of the nematode {\em Caenorhabditis elegans}:  neuro-sensory integration and propioceptive feedback}
\name{Charles Fieseler$^*$, James {Kunert-Graf}$^{**}$ and J. Nathan Kutz$^\dag$}
      
\address{$^*$Department of Physics, University of Washington, Seattle, WA 98195\\
$^{**}$ Pacific Northwest Research Institute, 720 Broadway, Seattle, WA 98122\\
$^\dag$ Department of Applied Mathematics, University of Washington, Seattle, WA 98195}

\begin{document}
\maketitle

\begin{abstract}
We develop a biophysically realistic model of the nematode {\em C. elegans} that includes:  (i) its muscle structure and activation, (ii) key connectomic activation circuitry, and (iii) a weighted and time-dynamic proprioception.  
In combination, we show that these model components can reproduce the {complex waveforms exhibited in {\em C. elegans} locomotive behaviors, chiefly omega turns}.
This is achieved via weighted, time-dependent suppression of the proprioceptive signal.
Though speculative, such dynamics are biologically plausible due to the presence of neuromodulators which have recently
been experimentally implicated in the escape response, which includes an omega turn.  
This is the first integrated neuromechanical model to reveal a mechanism capable of generating the complex waveforms observed in the behavior of {\em C. elegans}, thus contributing to a mathematical framework for understanding how control decisions can be executed at the connectome level in order to produce the full repertoire of observed behaviors. 
\end{abstract}


\section{Introduction}

Of general interest to the biology community is understanding how biomechanical systems process sensory input to produce behavioral outcomes and robust control strategies.  
Seemingly simple behavioral paradigms such as flying, crawling, and walking all involve complex interactions
between neuronal networks of sensory neurons, propioceptive feedback, and muscle activation. 
Understanding how these various networks interact to produce a robust control strategy remains an open challenge.
A model organism that can help elucidate the control laws arising from these complex dynamics is the {\em Caenorhabditis elegans}:  a nematode with only 302 neurons, 95 muscles involved in locomotion, and a well-mapped and stereotyped connectome~(\cite{White_paper,Varshney_connectome_update}).
Importantly, it has a limited behavioral repertoire that includes four primary motions:  forward crawling, backward crawling, and omega turns. 
In this manuscript, we explore a dynamic mechanism that can produce the full repertoire of turns in C elegans in a model optimized for forward motion.

Given its importance as a model organism, there { has long been an interest in modeling the behavior and locomotion of the worm (see (\cite{TheWholeWorm}) for a recent review). Broadly, these efforts (i) attempt to model the generation of locomotion within the nervous system alone (e.g. \cite{neuroModel1,neuroModel2,neuroModel3,James_paper,neuroModel4,neuroModel5,neuroModel6,neuroModel7}), (ii) model the biomechanics of the musculature/body alone (\cite{NewCohenModel,bodyModel1,bodyModel2,bodyModel3,bodyModel4,bodyModel5,bodyModel6,bodyModel7}), or (iii) build an integrated model for neural and bodily dynamics (\cite{short_SR,intModel1,intModel2,Biomech_gait,Dynamic_Neural_Network}).}
{Most previous modeling efforts have focused on simulating the simple, sinusoidal bodily postures involved in forward locomotion. It is unclear if said models could be extended to include the more complex behaviors exhibited by the worm.}
{Ultimately, the full complexity of the dynamics may be captured within future }
high-fidelity, fully three-dimensional {particle-based} models involving the collaboration of hundreds of scientists and modeling almost every aspect of the {\em C. elegans} geometry and anatomy~(\cite{Collaboration_models,Collaboration_models2}).
{To our knowledge the only model previously shown to be capable of generating complex postures is a non-integrated model of the body alone (\cite{NewCohenModel}). This model stops short of considering the role of neuronal dynamics and proprioception in generating complex postures.} 

{Nonetheless, integrated neuromechanical models have generated considerable insight into {\em C. elegans} locomotion. A notable recent example is the model of Boyle, Berri and Cohen~(\cite{Biomech_gait}), a two-dimensional spring-rod model which uses proprioception to generate sinusoidal locomotion.} The model incorporates proprioceptive feedback through specific stretch receptors, which have been long hypothesized~(\cite{Proprioception_hypothesis}), and for
which there is experimental evidence~(\cite{Proprioception_hypthothesis_TRP, Proprioception_hypothesis}).  {Via proprioceptive feedback, the model replicated the experimentally-observed continuous modulation of the worm's forward motion gait in response to its environment~(\cite{agar_drag}). However, this work considered only} forward motion, and {the model is unable to} reproduce other {typical behaviors} such as backward motion, head sweeps, or omega turns. 

{ In this manuscript, we extend the model of Boyle, Berri and Cohen~(\cite{Biomech_gait}), discovering the necessary modifications for the model to produce} the full range of {complex {\em C.elegans} postures}.
{Our modifications} produce a single biomechanically realistic model
that can produce the full repertoire of behaviors, including the ``omega turn'' in which the animal makes a deep bend in order
to reverse directions. We show that a traveling wave of suppression on the stretch
receptors is sufficient for this complex behavior.
This study suggests that transient, extra-synaptic
modulation of the synaptic weights is necessary for complex behavior, which is 
a vital step for understanding the control paradigm of the animal.

\section{Biomechanical Model}\label{sec:biomodel}

We review the two-dimensional spring-rod model ~(\cite{Biomech_gait}). This model integrates our dynamic proprioception which generates the repertoire of observed behaviors.

\subsection{Environmental properties}
This model implements the drag coefficient of the body by separating the parallel and
perpendicular components. In relatively low viscosity media similar
to water, the drag coefficients can be analytically calculated (\cite{Slender_body_theory}),
and in highly viscous media like agar, these coefficients have been
experimentally estimated~(\cite{agar_drag,short_SR}).
In the model, the medium is a linearly tunable parameter that varies from 0 (water) to 1 (agar), as shown in table 1.

\subsection{Model components}

\begin{figure*}[t!]
\centering
\begin{overpic}[width=1.0\textwidth]{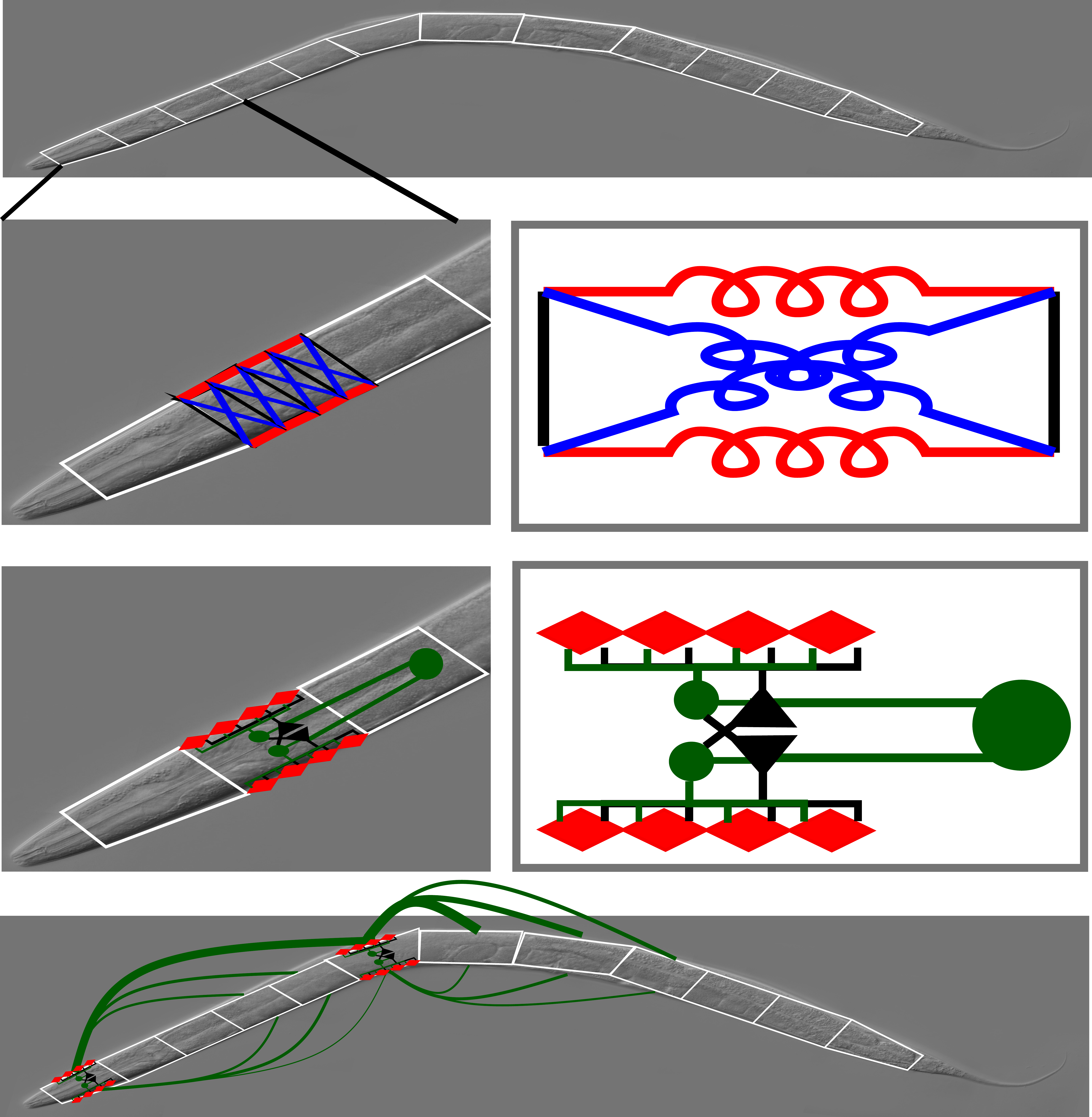}

\put(0.5,97){\color{white}\boxed{\text{a)}}}
\put(0.5,77){\color{white}\boxed{\text{b)}}}
\put(47,76.3){\boxed{\text{c)}}}
\put(0.5,46){\color{white}\boxed{\text{d)}}}
\put(47,46){\boxed{\text{e)}}}
\put(0.5,15){\color{white}\boxed{\text{f)}}}

\put(70,97){{\color{white} body segments}}
\put(76,96.5){{\color{white} \rotatebox{-115}{$\xrightarrow{\hspace*{0.3cm}}$} }}
\put(76,77.2){$L^{D}_{L,i}$}
\put(78,55){$L^{V}_{L,i}$}
\put(84,68){$L^{D}_{D,i}$}
\put(84,63.5){$L^{V}_{D,i}$}
\put(50,66){$R_i$}
\put(3,68){{\color{white} Rods \& springs}}
\put(4,37){{\color{white} Neural circuit}}
\put(49.7,42.3){\color{white}\text{VM}}
\put(57.1,42.3){\color{white}\text{VM}}
\put(64.5,42.3){\color{white}\text{VM}}
\put(71.9,42.3){\color{white}\text{VM}}

\put(60.5,36.2){\color{white}\text{VB}}
\put(66.5,35.3){\color{white}\text{VD}}

\put(49.7,24.8){\color{white}\text{DM}}
\put(57.1,24.8){\color{white}\text{DM}}
\put(64.6,24.8){\color{white}\text{DM}}
\put(71.9,24.8){\color{white}\text{DM}}

\put(60,31){\color{white}\text{DB}}
\put(66.5,32){\color{white}\text{DD}}

\put(89,34.5){\color{white}\text{AVB}}

\put(25,3){ \color{white} Proprioceptive signal during forward motion}
\put(30,2){\color{white}\rotatebox{-180}{$\xrightarrow{\linethickness{5pt}\hspace*{4cm}}$} }
\end{overpic}

\caption{Biomechanical model of C elegans. Based off of (\cite{Biomech_gait}).  a) The body has 12 segments. (b) and (c) Each segment has two rigid vertical components and four damped spring components. The diagonal (blue) elements are passive; the horizontal (red) elements are active and controlled by the neuron voltages. d,e) Each segment also has a simplified connectome model, with four pairs of ventral and dorsal motor neurons, and a pair of excitatory B-class neurons and inhibitory D-class neurons.  These are activated by a toy ``AVB-like'' command neuron, for forward motion. f) Proprioception produces oscillation and more complex behavior.  A small curvature will produce almost no proprioceptive signaling, but a stretched segment will. }
 \label{fig1}
%
\end{figure*}
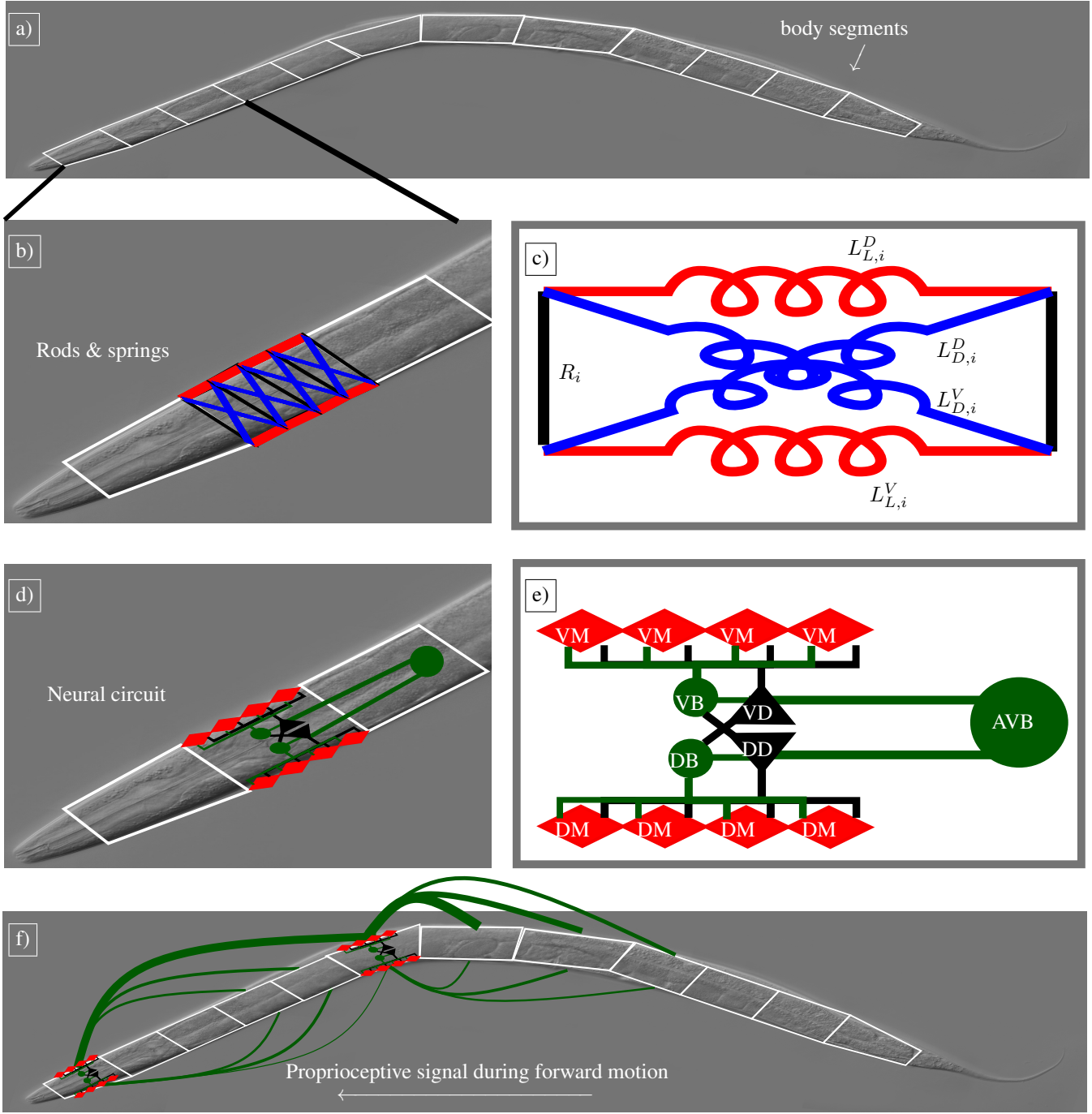

The two-dimensional model of the {\em C. elegans} has long been considered a compromise between
feasibility and accuracy~(\cite{2d_models}), i.e. it is a parsimonious model that balances complexity with accurate biomechanics.  The two-dimensional structure is motivated by the laboratory environment
where nearly the entire body moves only in two dimensions along a surface.  The only truly {three}-dimensional behaviors are exploratory head motions, which are outside the scope of this study.

\subsubsection{Body Shape and Segmentation}

The {\em C. elegans} body, as shown in Fig.~\ref{fig1}, is composed of
12 segments organized into 3 different layers of interaction. This
approximates the known muscle structure: {\em C. elegans} has 48 dorsal and 47 ventral
muscles, though the body itself is not segmented. A segment refers to 8 passive vertical
and diagonal elements containing a set of 4 dorsal and 4 ventral muscles, a
pair of stretch receptors, and a pair each of A- and B-class neurons.
The body is further divided into 48 sub-segments, 4 per full segment,
such that each has a pair of horizontal, diagonal, and vertical elements
and a single muscle.

The two-dimensional {cross-section} of the body is approximated by an ellipsoid, with the
radius of each of the $M=48$ sub-segments given by:
\begin{equation}
\text{R}_{i}=\text{R}_{0}\left|\sin\left(\arccos\left(\frac{i-\left(M/2+1\right)}{M/2+0.2}\right)\right)\right|\label{eq: Body radius}
\end{equation}
where $\text{R}_i$ is the radius of the $i$th body segment and $\text{R}_0$ is maximium radius.

\subsubsection{Rod spring model}

The first modeling component is a rod-spring system with passive vertical and
diagonal elements, as well as active muscle-driven horizontal elements.
The vertical rod elements are of a fixed length $2\text{R}_{i}$, given by
equation \ref{eq: Body radius}, and enforce the biological
constraint that the radius of the body is nearly constant throughout normal
behavior. 

The diagonal elements are damped springs that model hydrostatic internal
forces. The force from each diagonal element for the $i$th body segment is given by:
\begin{equation}
f_{D,i}^{k}=\kappa_{D}\left(L_{0D,i}-L_{D,i}^{k}\right)+\beta_{D}v_{D,i}^{k}
\end{equation}
where $\beta_D$ and $\kappa_D$ are the spring and damping constants,
and $L_{0D,i}=\sqrt{L_{seg}^{2}+\left(\text{R}_{i}+\text{R}_{i+1}\right)^{2}}$
is the rest length. In addition, $v_{i}=\frac{d}{dt}L_{i}^{k}$ is
the rate of change of the length of each element. The subscripts $D$
and, in the next equation, $L$, refer to either the diagonal or
horizontal elements. The superscript $k$ denotes which side of the animal
(dorsal or ventral) is being considered and which subnetwork is characterized (A-class or B-class).  
The four distinct values are thus $k=\{\text{A}, \text{B}\} \times \{\text{V}, \text{D}\}$ . 
Constants are identical across the subnetworks
unless otherwise noted. The horizontal elements are driven damped springs and the total force is
the sum of a passive and active component: $f_{Total,\ i}^{k} = f_{L,\ i}^{k} + f_{M,i}^{k} $.
The passive term is:
\begin{equation}
f_{H,\ i}^{k}=\begin{cases}
\kappa_{H}\left(L_{0H,i}\!-\!L_{H,i}^{k}\right)+\beta_{H}v_{H,i}^{k}\\
\ \ \ \ \ \ \ \ \ \ \ \ \mbox{for}\ \ L_{H,i}^{k}<L_{0H,i}\\
\kappa_{H}\left[\left(L_{0H,i}\!-\!L_{H,i}^{k}\right)+2\left(L_{0H,i}\!-\!L_{H,i}^{k}\right)^{4}\right]+\beta_{H}v_{H,i}^{k}\\
\ \ \ \ \ \ \ \ \ \ \ \  \mbox{otherwise}
\end{cases}
\end{equation}
The rest length for the horizontal elements is: $L_{0H,i}=\sqrt{L_{seg}^{2}+\left(\text{R}_{i}-\text{R}_{i+1}\right)^{2}}$.
The force output of a muscle segment is a function of the motor neuron
voltage, the muscle length, and the rate of contraction. In addition,
a gradient was imposed on the maximum output force of the muscles,
reflecting experimental observations
\begin{equation}
f_{M,i}^{k}=\kappa_{M,i}^{k}\left(L_{0M,i}^{k}-L_{H,i}^{k}\right)+\beta_{M,i}^{k}v_{H,i}^{k}
\end{equation}
with
\begin{eqnarray}
\kappa_{M,i}^{k} & \!\!\!=\!\!\! & \kappa_{0M}^{k}G^{k}_{\text{NMJ},i}\sigma\left(A_{M,i}^{k}\right) \\
L_{0M,i}^{k} & \!\!\!=\!\!\! & L_{0H,i}\!-\!\left(\Delta L\right)G^{k}_{\text{NMJ},i}\sigma\left(A_{M,i}^{k}\right) \\
\beta_{M,i}^{k} & \!\!\!=\!\!\! & \beta_{0M}\ G^{k}_{\text{NMJ},i}\sigma\left(A_{M,i}^{k}\right)
\end{eqnarray}
and where $\Delta L = L_{0H,i} - L_{\text{min,i}}$ is the rest length minus
a minimum muscle length for each subsegment, normalized
to have the same maximum curvature. $A_{M,i}^{k}$ is the muscle activation. The function $G^{k}_{\text{NMJ},i}$ is a linearly
decreasing function from the initiation of the propagation that captures the experimental fact
that curvature decreases as the wave propagates. Additionally, $\sigma\left(x\right)$
is a linearized sigmoidal function of the muscle activation:
\begin{equation}
\sigma\left(x\right)=\begin{cases}
0 & ,\ \ x\leq0\\
x & ,\ \ 0<x<1\\
1 & ,\ \ x\geq1
\end{cases}
\end{equation}

\begin{centering}
\begin{tabular}{|c|c|}
\hline 
Parameter & Value\tabularnewline
\hline 
\hline 
M & 48\tabularnewline
\hline 
N & 12\tabularnewline
\hline 
L & 1mm\tabularnewline
\hline 
$\text{L}_{\text{seg}}$ & L/M\tabularnewline
\hline 
$\text{CL}_{\text{water}}$ & $1.65\cdot10^{-6}/\left(\text{M}+1\right)$\tabularnewline
\hline 
$\text{CN}_{\text{water}}$ & $2.6\cdot10^{-6}/\left(\text{M}+1\right)$\tabularnewline
\hline 
$\text{CL}_{\text{agar}}$ & $1.6\cdot10^{-3}/\left(\text{M}+1\right)$\tabularnewline
\hline 
$\text{CN}_{\text{agar}}$ & $64\cdot10^{-3}/\left(\text{M}+1\right)$\tabularnewline
\hline 
$\kappa_{L}$ & $0.02$$\ \text{kg}\cdot\text{s}^{-1}$\tabularnewline
\hline 
$\kappa_{D}$ & $\kappa_{L}\cdot350$\tabularnewline
\hline 
$\kappa_{0M}$ & $\kappa_{L}\cdot20$\tabularnewline
\hline 
$\beta_{L}$ & $\kappa_{L}\cdot0.025\text{s}$\tabularnewline
\hline 
$\beta_{D}$ & $\kappa_{D}\cdot0.01\text{s}$\tabularnewline
\hline 
$\beta_{0M}$ & $\beta_{L}\cdot100$\tabularnewline
\hline 
$\text{L}_{0L,m}$ & $\sqrt{\text{L}_{seg}^{2}+\left(\text{R}_{m}-\text{R}_{m+1}\right)^{2}}$\tabularnewline
\hline 
$\text{L}_{0D,m}$ & $\sqrt{\text{L}_{seg}^{2}+\left(\text{R}_{m}+\text{R}_{m+1}\right)^{2}}$\tabularnewline
\hline 
$\Delta_{M}$ & $0.65\cdot\left(\text{R}_{m}+\text{R}_{m+1}\right)$\tabularnewline
\hline 
$\text{L}_{\text{min,m}}$ & $\text{L}_{0L,m}\frac{1-\Delta_{M}}{2R}$\tabularnewline
\hline 
$\text{R}_{0}$ & 40$\text{\ensuremath{\mu}m}$\tabularnewline
\hline 
$\epsilon_{\text{hyst}}$ & 0.5\tabularnewline
\hline 
\end{tabular}
\end{centering}

\subsubsection{Motor neurons}

A second critical component of the model is a simplified connectomic structure. In each segment,
the pair of 4 muscles receive input from two separate classes of excitatory
neurons. These A- and B-class motor neurons form separate subnetworks
that are experimentally well-known to be active in backwards and forwards
motion, respectively. Each neuron is modeled as bistable neuron which transitions 
instantanteously and is either ``on'' or ``off,'' $S=\{0,1\}$.
\begin{equation}
S_{i}^{k}=\begin{cases}
1 & \mbox{for} \ \ I_{i}^{k}>0.5+\epsilon_{hys}\left(0.5-S_{i}^{k}\right)\\
0 &  \mbox{otherwise}
\end{cases}
\end{equation}
Although there is some evidence that muscles display graded transmission
(\cite{Graded_synaptic_transmission}), there is also biological evidence
for bistability in the worm (\cite{Potential_bistability,Bistable_RMD, Action_potential_muscle}).
Previous work addressing this issue explicitly (\cite{Biomech_gait})
found no significant difference in behavior when the neurons
were modeled using a continuous model of the membrane voltage. The current
term is composed of three inputs into each of these motor neurons,
given by cross-inhibition, a ``command" neuron, and proprioception:
\begin{equation}
I_{i}^{k}=w_{-}^{k}S_{i}^{\bar{k}}+I_{\text{Command}}^{k}+I_{\text{SR},i}^{k}
\label{eq:input}
\end{equation}
The first two terms are explained in detail in the following paragraphs while
the third, {which contains the key contributions of this work, is} detailed in the next section.

In the real worm the contralateral inhibitory GABA-ergic D-class neurons synchronize muscle
contractions so that when one side is contracting the other is relaxing.
These D-class neurons are connected to the A-
and B-class neurons and their activity is highly correlated. Thus
in this model, cross-inhibition is applied directly in proportion
to the activity of the excitatory neurons on the opposite side, and
is captured in the term $w_{-}^{k}S_{i}^{\bar{k}}$. The second superscript,
$\bar{k}$, refers to the opposite side of the animal (dorsal or ventral).

In the full connectome, these motor neurons are part of larger locomotion
circuits and this is modeled here as the second input, from a single
{\em command neuron}.  This approximation does not assume that there 
exists a CPG for the production of oscillatory behavior, but is not incompatible
with a hybrid CPG and proprioceptive mechanism. Which (DC) current a neuron receives depends
on which subnetwork it is part of, with A-class (backwards) neurons
receiving current from the command ``AVA,'' and B-class (forwards) neurons
receiving current from command ``AVB.''

\begin{figure*}[t!]
\begin{centering}
\includegraphics[width=0.7\textwidth]{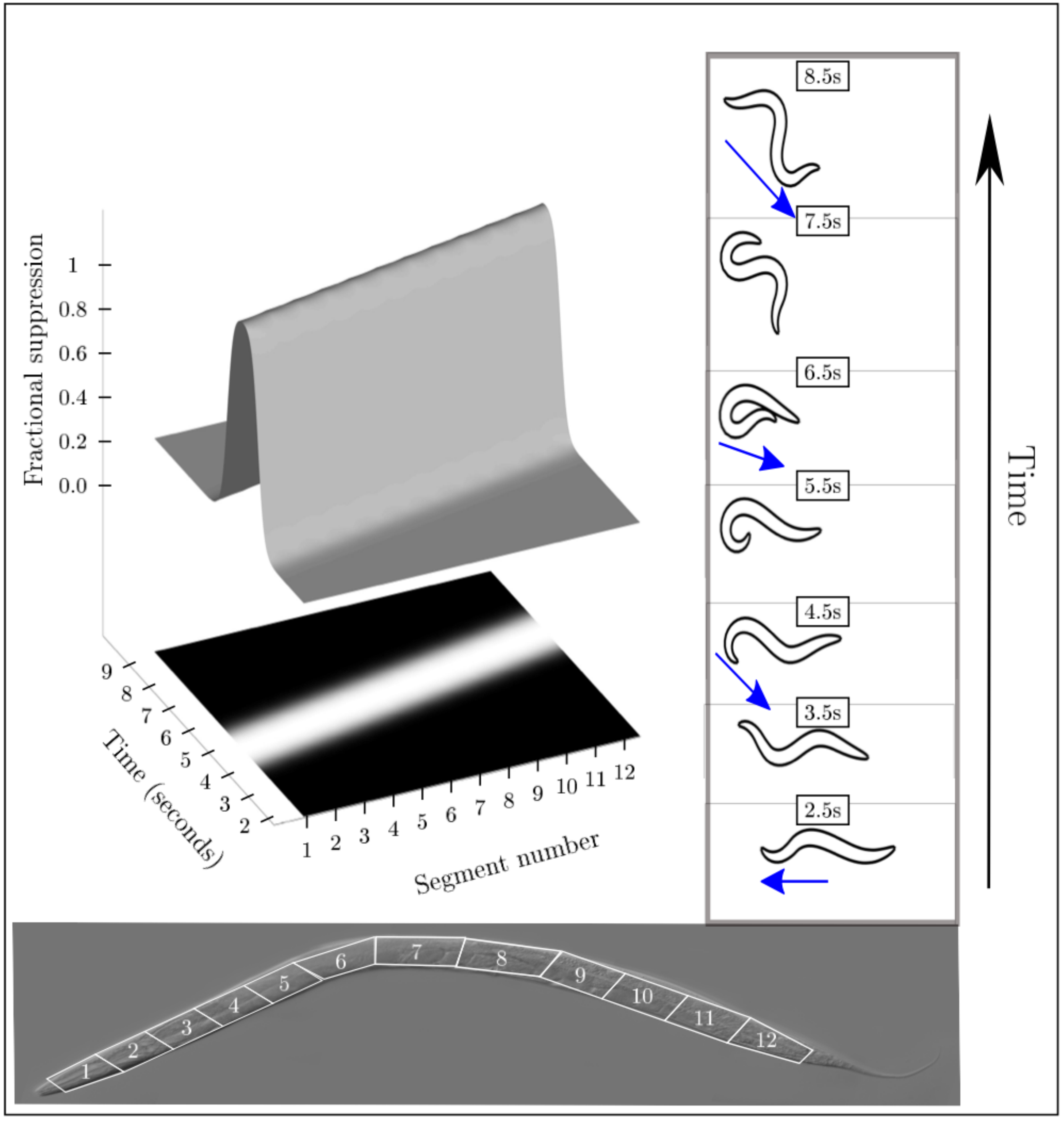}
\caption{
A wave of suppression on the stretch receptors produces an omega turn.
The percent suppressed is shown, which travels along lasting approximately
5 seconds.}
\label{fig:Wave of suppression}

\end{centering}
\end{figure*}

\section{Proprioception}\label{sec:proprio}

We now review the proprioceptive components of the model and introduce our modifications towards a more general dynamic model of proprioception.

\subsection{Stretch receptor current}

The remaining input (\ref{eq:input}) into the motor neurons, $I_{\text{SR},i}^{k}$, is also the
final component of the biomechanical model: proprioception. Stretch receptors
have long been hypothesized to exist due to long, undifferentiated
``arms'' that extend from the A- and B-class motorneurons down the
length of body (\cite{Proprioception_hypothesis}). Shown in Fig.~\ref{fig1} is how a stretched body segment produces a strong
signal for the posterior body segments on the same side, and a weak
to non-existent one on the opposite side. The number of segments to
be summed over is given by a parameter $s=\min \left(M;\ N_{\text{SR}}+(n-1)N_{out}\right)$,
which is a constant determined by the number of remaining posterior body segments.
The full sum is
\begin{equation}
I_{\text{SR},i}^{k}=\left(1-\alpha(t)\right) \cdot C_{i} \cdot G_{\text{SR},i}^{k}\sum_{j=(n-1)N_{out}+1}^{s}h_{j}^{k}
\label{eq:proprioception}
\end{equation}
where
\begin{equation}
C_{i}=\begin{cases}
1 & ,\ \ (n-1)N_{out}\leq M-N_{\text{SR}}\\
\sqrt{\frac{N_{\text{SR}}}{M-(n-1)N_{out}}} & ,\ \ (n-1)N_{out}>M-N_{\text{SR}}
\end{cases}.
\end{equation}
The term $(1-\alpha(t))$ is the time dependent term that allows for dynamic 
suppression of this current, and will be explained in the next section.
The parameter $A_i$ compensates for the fact that for segments close
to the posterior of the animal, there are fewer segment contributions.
Additionally, the parameter 
\begin{equation}
G_{\text{SR},i}^{k}=\begin{cases}
0.65\cdot\left(0.4+0.08\cdot\left(N-i-1\right)\cdot\frac{2N_{seg}}{12N_{seg\ per}}\right)\\
\ \ \ \ \ \ \ \ \ \ \ \ \mbox{for}\ \ k[0]=A\\
0.65\cdot\left(0.4+0.08\cdot i\cdot\frac{2N_{seg}}{12N_{seg\ per}}\right)\\
\ \ \ \ \ \ \ \ \ \ \ \ \mbox{for}\ \ k[0]=B
\end{cases}
\end{equation}
is a gradient that increases posteriorly for forward motion (B class)
and anteriorly for backward motion (A class), to make the receptors
more sensitive to the decreased curvature of the body (shown in figure
\ref{fig:Asymmetries}). The first element of $k$, $k[0]$, refers to the subnetwork. Finally, 
\begin{equation}
h_{i}^{k}=\lambda_{i}\gamma_{i}^{k}\frac{L_{H,i}^{k}-L_{0H,i}}{L_{0H,i}}
\end{equation}
is a stretch receptor activation function with parameters:
\begin{equation}
\lambda_{i}=\frac{2\text{R}_{0}}{\text{R}_{i}+\text{R}_{i+1}}
\end{equation}
which compensates for the variable radius of each segment with
\begin{equation}
\gamma_{i}^{k}=\begin{cases}
1 & ,\ \ k[1]=V\\
0.8 & ,\ \ k[1]=D;\ L_{H,i}^{k}>L_{0H,i}\\
1.2 & ,\ \ k[1]=D;\ L_{H,i}^{k}<L_{0H,i}
\end{cases}
\end{equation}
which compensates for the previously mentioned asymmetry in the inhibitory
circuit. The proprioceptive stretch sensors form the fundamental oscillatory
mechanism of the model.

An important note is that proprioceptive feedback for forward motion in our model can be described as an {\em anteriorly}
directed signal encouraging contraction from a {\em stretched} posterior segment, which
is consistent with the physiology of the B motor neurons (\cite{White_paper}). In contrast,
Quen et al. in (\cite{Proprioception_hypothesis}) provide experimental evidence that proprioception
acts as a {\em posteriorly} directed signal for contraction from a {\em contracted}
anterior segment. It is possible that both of these mechanisms are correct, and 
possible distinguishing experiments are discussed later.

\subsection{Dynamics of proprioception}
Unlike simple forward and backward locomotion, which are long-lived oscillations
of the network, the omega turn is a transient behavior which only
lasts a few seconds. We phenomenologically model this
as a wave of modulation in the proprioceptive signal
that travels posteriorly along the body. 
Sigmoidal functions are common in biological systems, so we 
posit a two-sided function:
\begin{align}
\alpha (t) =\frac{1}{2} \left[\tanh\left(s\left(t-t_{start}\right)\right)-\tanh\left(s\left(t-t_{end}\right)\right)\right]
\end{align}
where $t_{start}$ and $t_{end}$ are respectively the initiation and completion of wave,
 and $s$ models the speed at which this suppression takes effect. 
This function can smoothly tune a parameter to 0 and then
back to its full value, as well as increase or decrease parameters
by a percentage using $1\pm\alpha$. This addition to the original model is vital for complex
and transient behaviors like the omega turn and other shallow turns, and is implemented in equation \ref{eq:proprioception}.

\subsection{Numerical modeling}

The original model used Sundials version 2.3.0 (\cite{Biomech_gait}); this paper
uses version 2.6.1. The numerical simulation portion of the
code is written in C++. Based on the original paper, a visualization
package written in MATLAB 2016b is included. The model and dynamics proposed here are
all fully reproducible, with the code and example datasets available at (\cite{GitHub}).

\section{Results}

We show our model can produce omega turns within the model using modulated proprioception.

\subsection{Backwards motion}

\begin{figure}[t!]
\includegraphics[width=0.45\textwidth]{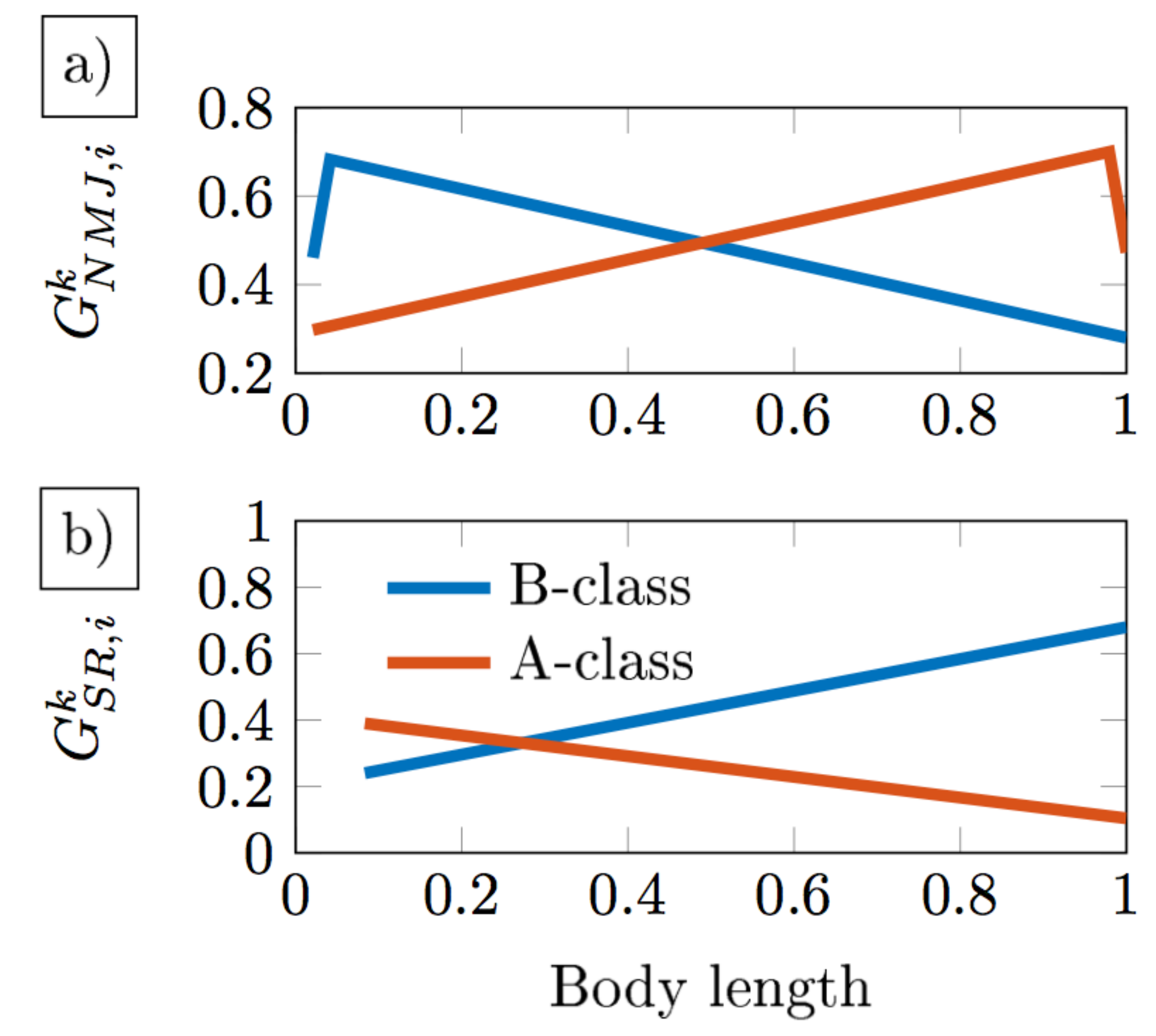}
\caption{
Asymmetries needed to {produce} backwards and forward motion. a) The
nueromuscular junctions (NMJs) decrease in strength as you travel
posteriorly (anteriorly) along the body for forward (backward) motion
and B- (A-) class neurons. The head (tail) is weakened in the original
model in order to produce straight forward motion, and there is recent
experimental evidence that the head circuit is fine tuned in a similar
way (\cite{Head_angle_fine_tuning}). b) Partially to compensate for
the decrease in NMJ strength, the stretch receptor sensitivity is
increased as you travel posteriorly (anteriorly) along the body for
forward (backward) motion. }
\label{fig:Asymmetries}
\end{figure}

There are three front-to-back asymmetries that bias the original
model towards forwards motion. Two of them are shown in Fig.~\ref{fig:Asymmetries}.
For forward motion they are: a decrease in muscle strength along
the length of the body, and an increase in stretch sensitivity to
partially compensate for this. 

Importantly, the A- and B-class subnetworks of neurons have ``arms''
extending in opposite directions down the length of the body (\cite{White_paper}), and
this is modeled as a stretch producing a signal in the anterior body
segments for the A neurons. In this way, backwards motion can be plausibly
and simply modeled using a mirrored subnetwork of motor neurons with
oppositely aligned stretch receptors. 

\subsection{Optimized stretch receptors}


\begin{figure}[t!]
\includegraphics[width=0.45\textwidth]{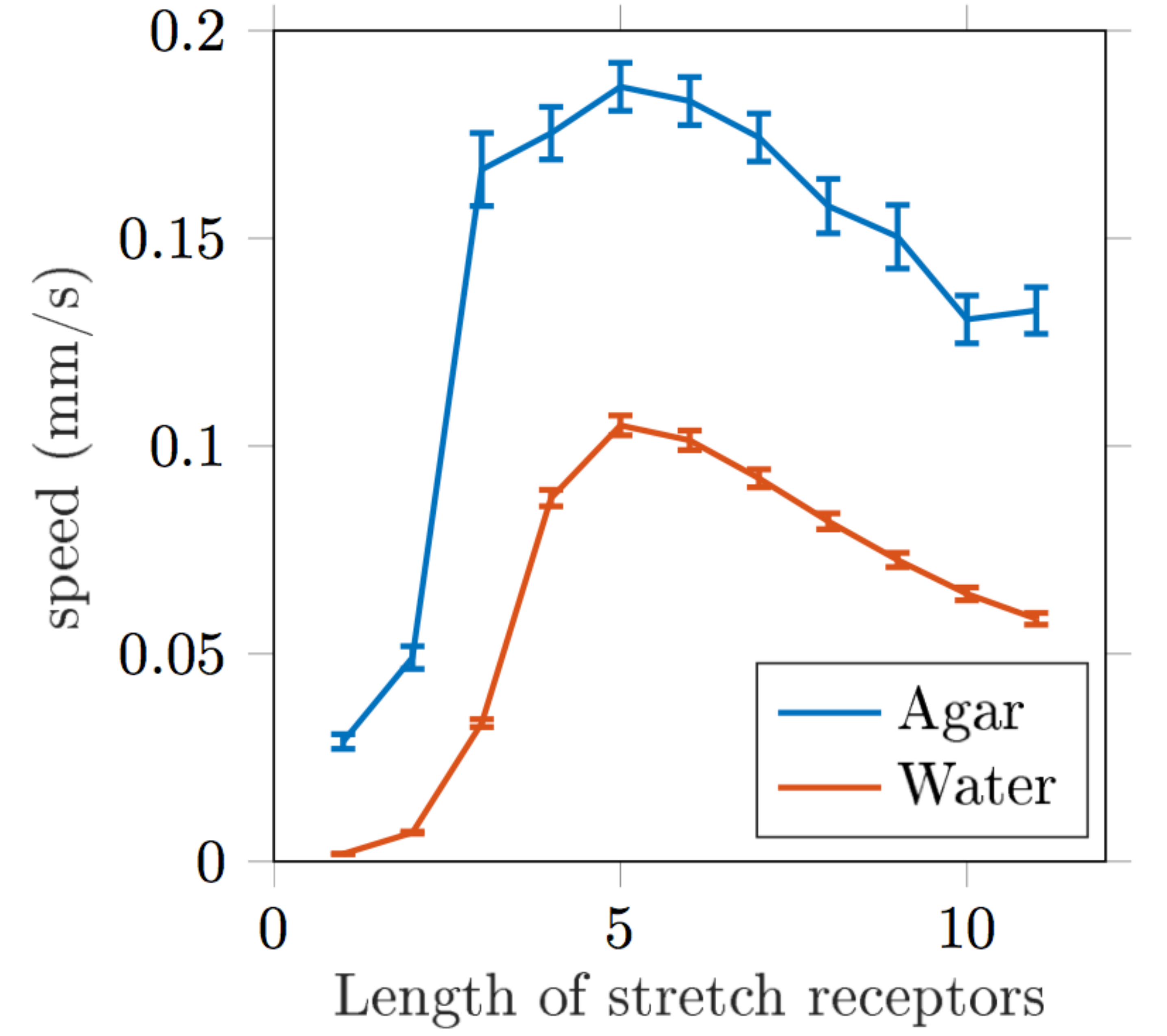}
\caption{
Average Center Of Mass velocity for regular forward motion as a function
of the length of the stretch receptors, measured in body segments.}
\label{fig:VCoM}
\end{figure}

Proprioceptive receptors have long been postulated, but though there
are very suggestive experiments (\cite{Proprioception_hypothesis,Proprioception_hypthothesis_TRP,Proprioceptive_hypothesis_other_neurons}),
it is not known through what mechanism the worm senses stretching.
Physiological data reveals the presence of long undifferentiated ``arms''
that stretch away from the B- and A-class motor neurons for approximately a quarter
of the body length, but their function is unknown. To contribute
to constraints on this hypothesis, which is necessary for both simple
and complex behaviors in this model, studies on the effect of changing
the length of the body receptors on speed of forward motion and other
metrics were performed. Figure \ref{fig:VCoM} shows that there is
a maximum center of mass velocity for proprioceptive sensors of length
equal to approximately 5-6 segments, which would allow each segment
to receive information about one half body wavelength in agar. A
pronounced feature of figure \ref{fig:VCoM} is the drastic decrease
in speed for very short stretch receptors of approximately 1-2
segment lengths.

\subsection{Traveling waves of suppression produce omega turns}

The addition of a set of parameters that controls a wave of suppression
on the stretch receptors along the body wall is enough to realistically
produce an omega turn in the model. Figure \ref{fig:Wave of suppression}
shows a ventral turn, though this mechanism can produce turns to either
side.

To understand this behavior, it is instructive to understand
what happens to the different body segments as the wave passes through
them. When the wave is initialized in the head segment (segment 1),
the head becomes less able to sense the curvature of the segments
immediately posterior. As the wave travels across the next few 
segments, the first third of the body continues to tighten its
turn because the proprioceptive input is no longer present to stimulate
the dorsal muscles, those opposite to those currently active. This
tightening continues until the wave passes and the first segments
slowly become able to sense the extreme curvature of the first half
of the body. The head starts to unwind, producing a smooth change in the direction of
motion. By tuning the timescale or level of suppression of this wave, the mechanism is able
to produce turns of any angle.

The continued suppression of the stretch receptors on the posterior
half of the body once the head has started to resume normal forward
motion is vital to the success of this maneuver. In a realistic omega
turn the rest of the body follows the head through the highly curved
``omega'' shape. In this model, deep bending is resisted by any part of the
body in which the proprioceptive signals remain unchanged. Non-traveling suppression
of stretch receptors in one part of the body produces various types
of thrashing behavior, paralysis, or changes in gait. Furthermore,
the forward momentum through the curving head bend must be produced
by the continued undulations of the posterior half of the body. Thus,
if the proprioceptive hypothesis of the Cohen model is correct, a traveling 
silencing of proprioception is a simple way to produce this complex behavior.

\subsection{Robustness of omega turns}


\begin{figure}[t!]
\includegraphics[width=0.45\textwidth]{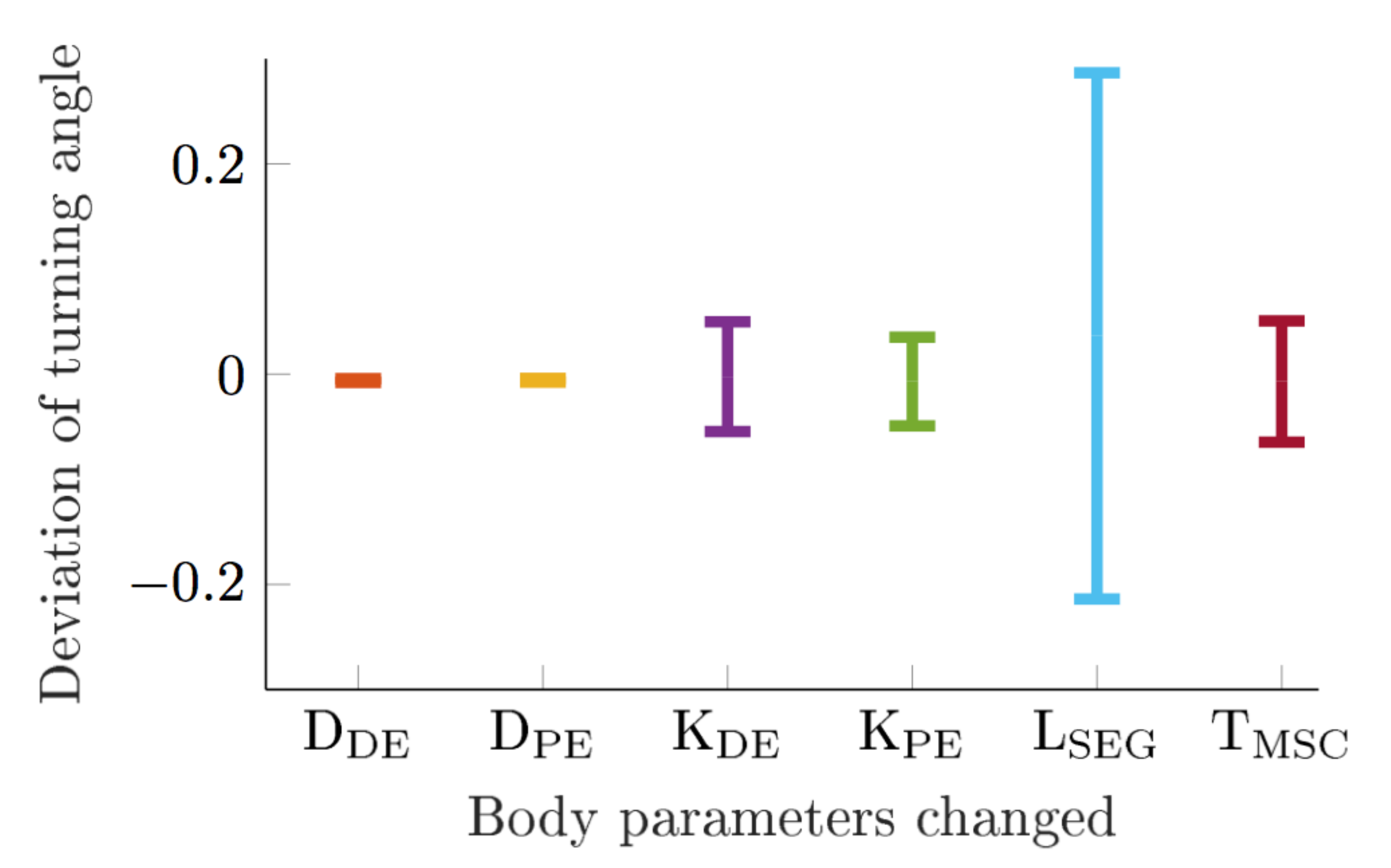}
\caption{
Angle change as a function of various body parameters. Displayed is
the mean and standard deviation when the simulation is run for individuals
with up to 10\% variation in these parameters: $D_{*}$= damping coefficient;
$k_{*}$= spring constant; $*_{PE}$= horizontal (passive) element;
$*_{DE}$= diagonal element; $L_{SEG}$= segment length; $T_{MSC}$=
muscle time constant. }
\label{fig:Ensemble of models}
\end{figure}

As discussed in (\cite{Generic_modeling}), a model of a complex system
that uses average values of parameters, as this model does, often
gives non-average results. Therefore, it is paramount to demonstrate
this behavior in an ensemble of models with different internal parameters,
and a sensitivity analysis is shown in Fig.~\ref{fig:Ensemble of models}.
Though the model is relatively sensitive to overall body length,
even in this case the turning angle only changed about 0.2 radians, thus suggesting
the mechanism can robustly produce omega turns.

\section{Discussion}

We have presented a biophysically realistic model that can reproduce
the essential repertoire of {\em C. elegans} locomotive behaviors: forward
motion, backward motion, and omega turns. Recent work has shown that
extra-synaptic modulation is important in more complex behavioral
patterns in {\em C. elegans} (\cite{Tyramine_paper}), and this is the first
biomechanical model to our knowledge that uses this information and
proposes an interpretable mechanism for behaviors beyond forward motion.

This model relies fundamentally on the hypothesis that there exist
stretch receptors in {\em C. elegans}, and that they are posteriorly 
(anteriorly) directed for B(A)-class neurons. The model further allows for 
testable predictions about the characteristics of
those receptors. Mutant studies should be
able to identify potential chemical or neural candidates that can control
proprioception, which in turn might
help illuminate the network involved in proprioception. Another class
of experimental tests is optogenetic manipulation of worms 
trapped in microfluidic devices along the lines of Quen et al. (\cite{Proprioception_hypothesis}).
If neurons associated with omega turns, e.g. RIV, SMDV, or RIM, are 
stimulated and silencing of the proprioceptive
signal is measured, that would be strong evidence for this mechanism.

A traveling wave of silencing of the proprioceptive
stretch receptors can robustly produce omega turns, but
we also found that an increase in muscle strength can drastically
increase the turning angles of the worm. Though it is possible that
the muscle activation is modulated during this behavior, quantitative 
statements are complicated by many approximations. These issues are
discussed in more depth in the original model paper (\cite{Biomech_gait}).
Thus, unlike the qualitative proposal of proprioceptive modulation,
 no quantitative statement can be made
about muscle dynamics.

A key limitation of this work is the biological plausibility of the traveling wave of 
suppression. A derivation from first principles is left for future work, but a
promising line of research is nonlinear diffusion equations with
traveling wave solutions  
(\cite{chen2016existence, hosono1987traveling, atkinson1981traveling, wu2001traveling}).
Another direction is to study a diffusive neuromodulator in combination with
the underlying synaptic network (\cite{bentley2016multilayer}), 
as a wave-like effect might be produced by this interplay. 
This modeling work can lend intuition to future work,
which may give more support to an alternate mathematical form for this wave.

Future work could also explore even more complex behavior.
Our mechanism can produce shallower or deeper
turns, both of which have been proposed as distinct categories of
behavior (\cite{Shallow_turn,Coil_turn}). The ``jockeying'' of a worm interrupted 
during an omega turn could be explored via the release of multiple waves of silencing
in succession. 
In future work, we hope to integrate the connectomic dynamics
with the proposed biomechanial model in order to understand the "inside" and "outside" of the
worm and how the connectome serves as the controller for behavioral outputs.

Much recent modeling work has contributed to discussions surrounding
simple behaviors like forward motion, and we hope that this paper
can contribute to similar discussions of more complex behavioral dynamics. 
The ability of a single model to produce basic motion
and an omega turn helps elucidate the control structure of {\em C. elegans}, and
can help inform what types of outputs should be produced by the internal
dynamics of the network~(\cite{James_paper}). 
Importantly, we have illustrated that dynamic processes can
play a critical role in controlling the {\em C. elegans} repertoire of behavior.

\section*{Conflict of Interest}

The authors declare no conflict of interest associated with this paper.

\section*{Acknowledgements}

We are indebted to Aravi Samuel for help comments and insights into the {\em C. elegans} locomotion and dynamics.
C. Fieseler acknowledges support from a National Science Foundation Graduate Research Fellowship under Grant No. DGE-1256082.

\section*{References}
\bibliography{celegan_biomech_arxiv2}{}

\begin{thebibliography}{10}

\bibitem{Proprioceptive_hypothesis_other_neurons}
A.~Albeg, C.~J. Smith, M.~Chatzigeorgiou, D.~G. Feitelson, D.~H. Hall, W.~R.
  Schafer, D.~M. Miller, and M.~Treinin.
\newblock C. elegans multi-dendritic sensory neurons: morphology and function.
\newblock {\em Molecular and Cellular Neuroscience}, 46(1):308--317, 2011.

\bibitem{atkinson1981traveling}
C~Atkinson, GEH Reuter, and CJ~Ridler-Rowe.
\newblock Traveling wave solution for some nonlinear diffusion equations.
\newblock {\em SIAM Journal on Mathematical Analysis}, 12(6):880--892, 1981.

\bibitem{bodyModel6}
M.~Backholm, A.~K.~S. Kasper, R.~D. Schulman, W.~S. Ryu, and K.~Dalnoki-Veress.
\newblock The effects of viscosity on the undulatory swimming dynamics of c.
  elegans.
\newblock {\em Phys Fluids}, 27:091901, 2015.

\bibitem{bentley2016multilayer}
Barry Bentley, Robyn Branicky, Christopher~L Barnes, Yee~Lian Chew, Eviatar
  Yemini, Edward~T Bullmore, Petra~E V{\'e}rtes, and William~R Schafer.
\newblock The multilayer connectome of caenorhabditis elegans.
\newblock {\em PLoS computational biology}, 12(12):e1005283, 2016.

\bibitem{agar_drag}
S.~Berri, J.~H. Boyle, M.~Tassieri, I.~A. Hope, and N.~Cohen.
\newblock Forward locomotion of the nematode c. elegans is achieved through
  modulation of a single gait.
\newblock {\em HFSP journal}, 3(3):186--193, 2009.

\bibitem{Collaboration_models}
A.~Blau, F.~Callaly, S.~Cawley, A.~Coffey, A.~De~Mauro, G.~Epelde, L.~Ferrara,
  F.~Krewer, C.~Liberale, P.~Machado, and G.~Maclair.
\newblock July.
\newblock In In~Conference on~Biomimetic and .~Biohybrid~Systems, editors, {\em
  The Si elegans Project\textendash The Challenges and Prospects of Emulating
  Caenorhabditis elegans}, pages 436--438, International Publishing, 2014.
  Springer.

\bibitem{Biomech_gait}
J.~H. Boyle, S.~Berri, and N.~Cohen.
\newblock Gait modulation in c. elegans: an integrated neuromechanical model.
\newblock {\em Frontiers in computational neuroscience}, 6:10, 2012.

\bibitem{intModel2}
J.~H. Boyle, J.~Bryden, and N.~Cohen.
\newblock An integrated neuro-mechanical model of c. elegans forward
  locomotion.
\newblock {\em Lect Notes Comput Sci}, 4984:37--47, 2008.

\bibitem{Coil_turn}
O.~D. et~al. Broekmans.
\newblock Resolving coiled shapes reveals new reorientation behaviors in c.
  elegans.
\newblock {\em eLife}, 5, 2016.

\bibitem{intModel1}
J.~A. Bryden and N.~Cohen.
\newblock Neural control of caenorhabditis elegans forward locomotion: the role
  of sensory feedback.
\newblock {\em Biol Cybern}, 98:339--351, 2008.

\bibitem{chen2016existence}
Xinfu Chen, Yuanwei Qi, and Yajing Zhang.
\newblock Existence of traveling waves of auto-catalytic systems with decay.
\newblock {\em Journal of Differential Equations}, 260(11):7982--7999, 2016.

\bibitem{NewCohenModel}
N.~Cohen and T.~A Ranner.
\newblock {\em New Computational Method for a Model of C. elegans Biomechanics:
  Insights into Elasticity and Locomotion Performance.}
\newblock [physics.bio-ph], 2017.

\bibitem{Generic_modeling}
D.~D. Cook and D.~J. Robertson.
\newblock The generic modeling fallacy: Average biomechanical models often
  produce non-average results{!}
\newblock {\em Journal of Biomechanics}, 49(15):3609--3615, 2016.

\bibitem{Dynamic_Neural_Network}
X.~Deng, J.~X. Xu, J.~Wang, G.~Y. Wang, and Q.~S. Chen.
\newblock Biological modeling the undulatory locomotion of c. elegans using
  dynamic neural network approach.
\newblock {\em Neurocomputing}, 186:207--217, 2016.

\bibitem{Tyramine_paper}
J.~L. Donnelly, C.~M. Clark, A.~M. Leifer, J.~K. Pirri, M.~Haburcak, M.~M.
  Francis, A.~D. Samuel, and M.~J. Alkema.
\newblock Monoaminergic orchestration of motor programs in a complex c. elegans
  behavior.
\newblock {\em PLoS Biol}, 11:4, 2013.

\bibitem{bodyModel1}
P.~{Erd\"os} and E.~Niebur.
\newblock The neural basis of the locomotion of nematodes.
\newblock {\em Lect Notes Phys}, 1990(368):253--267, 1990.

\bibitem{GitHub}
Charles Fieseler.
\newblock Github.
\newblock Code in MATLAB and C++, 2017.

\bibitem{Action_potential_muscle}
S.~Gao and M.~Zhen.
\newblock Action potentials drive body wall muscle contractions in
  caenorhabditis elegans.
\newblock {\em Proceedings of the National Academy of Sciences},
  108(6):2557--2562, 2011.

\bibitem{Potential_bistability}
A.~et~al. Gordus.
\newblock Feedback from network states generates variability in a probabilistic
  olfactory circuit.
\newblock {\em Cell}, 161(2):215--227, 2015.

\bibitem{hosono1987traveling}
Yuzo Hosono.
\newblock Traveling waves for some biological systems with density dependent
  diffusion.
\newblock {\em Japan Journal of Applied Mathematics}, 4(2):297, 1987.

\bibitem{2d_models}
E.~J. Izquierdo and R.~D. Beer.
\newblock An integrated neuromechanical model of steering in c. elegans.
\newblock In {\em Proceeding of the European Conference on Artificial Life},
  pages 199--206, 2015.

\bibitem{TheWholeWorm}
E.~J. Izquierdo and R.~D. Beer.
\newblock The whole worm: brain-body-environment models of c. elegans.
\newblock {\em Current Opinion in Neurobiology}, 40:23--30, 2016.

\bibitem{neuroModel3}
J.~Karbowski, G.~Schindelman, C.~J. Cronin, A.~Seah, and P.~W. Sternberg.
\newblock Systems level circuit model of c. elegans undulatory locomotion:
  mathematical modeling and molecular genetics.
\newblock {\em J Comput Neurosci}, 24:253--276, 2008.

\bibitem{Shallow_turn}
D.~Kim, S.~Park, L.~Mahadevan, and J.~H. Shin.
\newblock The shallow turn of a worm.
\newblock {\em Journal of Experimental Biology}, 214(9):1554--1559, 2011.

\bibitem{James_paper}
J.~Kunert, E.~Shlizerman, and J.~N. Kutz.
\newblock Low-dimensional functionality of complex network dynamics:
  Neurosensory integration in the caenorhabditis elegans connectome.
\newblock {\em Physical Review E}, 89:5, 2014.

\bibitem{neuroModel6}
J.~M. Kunert, P.~D. Maia, and J.~N. Kutz.
\newblock Functionality and robustness of injured connectomic dynamics in c.
  elegans: Linking behavioral deficits to neural circuit damage.
\newblock {\em PLoS Computational Biology}, 13:1, 2017.

\bibitem{neuroModel5}
J.~M. Kunert, J.~L Proctor, S.~L Brunton, and J.~N. Kutz.
\newblock Spatiotemporal feedback and network structure drive and encode
  caenorhabditis elegans locomotion.
\newblock {\em PLoS Computational Biology}, 13:1, 2017.

\bibitem{neuroModel7}
J.~M. Kunert-Graf, E.~Shlizerman, and J.~N. Kutz.
\newblock Multistability and long-timescale transients encoded by network
  structure in a model of c. elegans connectome dynamics.
\newblock {\em Frontiers in Computational Neuroscience}, 11, 2017.

\bibitem{bodyModel7}
S.~H. Lee and Kang Sh.
\newblock Characterization of the crawling activity of caenorhabditis elegans
  using a hidden {M}arkov model.
\newblock {\em Theory Biosci}, 134:117--125, 2015.

\bibitem{Proprioception_hypthothesis_TRP}
W.~Li, Z.~Feng, P.~W. Sternberg, and X.~S. Xu.
\newblock A c. elegans stretch receptor neuron revealed by a mechanosensitive
  trp channel homologue.
\newblock {\em Nature}, 440(7084):684--687, 2006.

\bibitem{Slender_body_theory}
J.~Lighthill.
\newblock Flagellar hydrodynamics.
\newblock {\em SIAM review}, 18(2):161--230, 1976.

\bibitem{Graded_synaptic_transmission}
Q.~Liu, G.~Hollopeter, and E.~M. Jorgensen.
\newblock Graded synaptic transmission at the caenorhabditis elegans
  neuromuscular junction.
\newblock {\em Proceedings of the National Academy of Sciences},
  106(26):10823--10828, 2009.

\bibitem{bodyModel3}
R.~Mailler, J.~Avery, J.~Graves, and N.~Wily.
\newblock Biologically accurate 3d model of the locomotion of caernorhabditis
  elegans.
\newblock In {\em International Conference on Biosciences}, pages 84--90, 2010.

\bibitem{bodyModel4}
T.~Majmudar, E.~E. Keaveny, J.~Zhang, and M.~J. Shelley.
\newblock Experiments and theory of undulatory locomotion in a simple
  structured medium.
\newblock {\em J R Soc Interface}, 9:1809--1823, 2012.

\bibitem{Bistable_RMD}
J.~E. Mellem, P.~J. Brockie, D.~M. Madsen, and A.~V. Maricq.
\newblock Action potentials contribute to neuronal signaling in c. elegans.
\newblock {\em Nature neuroscience}, 11(8):865--867, 2008.

\bibitem{neuroModel1}
E.~Niebur and P.~{Erd\"os}.
\newblock {\em Computer simulation of networks of electrotonic neurons}.
\newblock Cambridge University Press, In Computer Simulation in Brain Science.
  . pp.148-163, 1988.

\bibitem{short_SR}
E.~Niebur and P.~{Erd\"os}.
\newblock Theory of the locomotion of nematodes: dynamics of undulatory
  progression on a surface.
\newblock {\em Biophysical journal}, 60(5):1132--1146, 1991.

\bibitem{neuroModel4}
T.~E. Portegys.
\newblock Training sensory-motor behavior in the connectome of an artificial c.
  elegans.
\newblock {\em Neurocomputing}, 168:128--134, 2015.

\bibitem{bodyModel5}
Y.~Rabets, M.~Backholm, K.~Dalnoki-Veress, and W.~S. Ryu.
\newblock Direct measurements of drag forces in c. elegans crawling locomotion.
\newblock {\em Biophys J}, 107:1980--1987, 2014.

\bibitem{bodyModel2}
M.~{R\"onkk\"o} and G.~Wong.
\newblock Modeling the c. elegans nematode and its environment using a particle
  system.
\newblock {\em J Theor Biol}, 253:316--322, 2008.

\bibitem{neuroModel2}
K.~Sakata and R.~Shingai.
\newblock Neural network model to generate head swing in locomotion of
  caenorhabditis elegans.
\newblock {\em Network}, 15:199--216, 2004.

\bibitem{Head_angle_fine_tuning}
Y.~et~al. Shen.
\newblock An extrasynaptic gabaergic signal modulates a pattern of forward
  movement in caenorhabditis elegans.
\newblock {\em eLife}, 5, 2016.

\bibitem{Collaboration_models2}
B.~Szigeti, P.~Gleeson, M.~Vella, S.~Khayrulin, A.~Palyanov, J.~Hokanson,
  M.~Currie, M.~Cantarelli, G.~Idili, and S.~Larson.
\newblock Openworm: an open-science approach to modeling caenorhabditis
  elegans.
\newblock {\em Frontiers in computational neuroscience}, 8:137, 2014.

\bibitem{Varshney_connectome_update}
L.~R. Varshney, B.~L. Chen, E.~Paniagua, D.~H. Hall, and D.~B. Chklovskii.
\newblock Structural properties of the caenorhabditis elegans neuronal network.
\newblock {\em PLoS Comput Biol}, 7:2, 2011.

\bibitem{Proprioception_hypothesis}
Q.~Wen, M.~D. Po, E.~Hulme, S.~Chen, X.~Liu, S.~W. Kwok, M.~Gershow, A.~M.
  Leifer, V.~Butler, C.~Fang-Yen, and T.~Kawano.
\newblock Proprioceptive coupling within motor neurons drives c. elegans
  forward locomotion.
\newblock {\em Neuron}, 76(4):750--761, 2012.

\bibitem{White_paper}
J.~G. White, E.~Southgate, J.~N. Thomson, and S.~Brenner.
\newblock The structure of the nervous system of the nematode caenorhabditis
  elegans.
\newblock {\em Philos Trans R Soc Lond B Biol Sci}, 314(1165):1--340, 1986.

\bibitem{wu2001traveling}
Jianhong Wu and Xingfu Zou.
\newblock Traveling wave fronts of reaction-diffusion systems with delay.
\newblock {\em Journal of Dynamics and Differential Equations}, 13(3):651--687,
  2001.

\end{thebibliography}
\bibliographystyle{plain}

\end{document}